\documentstyle[prb,multicol,aps,epsfig]{revtex}

\begin{document}

\title{Half-metallicity and efficient spin injection 
in AlN/GaN:Cr (0001) heterostructure
}

\author{J.~E. Medvedeva$^1$, A.~J. Freeman$^1$, X.~Y. Cui $^2$, C. Stampfl$^2$ 
and N. Newman$^3$
}

\address{
$^1$ Physics and Astronomy Department, Northwestern University, 
Evanston, Illinois 60208-3112 \\
$^2$ School of Physics, University of Sydney, Sydney, Australia\\
$^3$ Chemical and Materials Engineering Department,
Arizona State University, Tempe, Arizona 85287-6006
}

\maketitle

\begin{abstract}
First-principles investigations of the structural, electronic and magnetic properties 
of Cr-doped AlN/GaN (0001) heterostructures reveal that Cr segregates into 
the GaN region, that these interfaces retain their important half-metallic 
character and thus yield efficient (100 \%) spin polarized injection 
from a ferromagnetic GaN:Cr electrode through an AlN tunnel barrier
-- whose height and width can be controlled by adjusting the 
Al concentration in the graded bandgap engineered Al$_{1-x}$Ga$_x$N (0001) layers. 
\end{abstract}

\vspace{0.2cm}

\begin{multicols}{2}

The injection and manipulation of spin polarized carriers --
a key feature determining successful spintronics -- is highly dependent
on phenomena and proccesses occuring at the interface between a
ferromagnet and a semiconductor. 
It has been shown \cite{Soulen,Ristoiu,Groot} that surface or interface 
sensitive approaches must be employed in both theory and experiment for 
the accurate determination of the spin polarization in materials proposed for 
potential magnetoelectronic devices.
As perhaps the most striking example, Heusler compounds (such as NiMnSb, Co$_2$MnGe,
Co$_2$CrAl) -- which possess the appealing half-metallic ferromagnetic behavior 
in bulk -- show significantly reduced spin polarization at the surface 
\cite{Ristoiu} or, consequently, at the interface with a semiconductor 
\cite{Groot}.
Indeed, first-principles calculations have revealed that structural reconstruction
in the vicinity of the junction essentially alters
the electronic states and destroys half-metallicity \cite{Groot}.

Recently, nitride-based semiconductors, in particular GaN and AlN, 
have attracted increasing attention for a number of reasons: 
(i) The introduction of magnetic dopants and achieving ferromagnetism in these 
materials provide complementary functionality to the wide range of devices 
already developed for pure large band-gap nitrides \cite{books}. 
(ii) The shorter bond length and smaller spin-orbit coupling in these 
light-element compounds as compared to other III-V semiconductors (e.g., GaAs) 
are predicted to give rise to higher Curie temperatures \cite{Dietl00}; indeed,
recent measurements in Cr-doped GaN and AlN bulk
showed T$_c$ to be over 900~K \cite{Newman}.
(iii) The smaller spin-orbit interaction is also one of the main arguments 
for longer (by three orders of magnitude) electron spin lifetimes in GaN 
than in GaAs \cite{lifetimes}. 
(iv) The III-N semiconductors appear to be much more resilient to the 
presence of extended structural defects than are other semiconductors 
\cite{Lester95}.
While Cr-doped bulk GaN and AlN systems have been extensively studied 
\cite{Newman,Mryasov,Park,Das}, 
both experimental investigations and theoretical modeling 
of magnetically doped III-N interfaces -- essential
for practical spintronic applications -- are lacking.

In this Letter, we present results of first-principles calculations of
Cr-doped AlN/GaN (0001) heterostructures focusing on their structural, 
electronic and magnetic properties. 
From a comparison of the formation energy of the relaxed AlN/GaN 
heterostructures with different
site locations of Cr, we predict that the magnetic impurity segregates
into the GaN region which serves as a ferromagnet, while
AlN is the semiconductor part of the interface. 
Most significantly, we find that the Cr-doped nitride-based interface
retains the desired half-metallic behavior that enables  
the efficient 100 \% spin polarized injection across the interface and thus 
makes this system highly attractive for spintronic applications.
Finally, we propose a way to control the shape (i.e., height and width)
of the nitride potential barrier by adjusting the concentration $x$ in a 
Cr-doped interface with a graded Al$_{1-x}$Ga$_x$N barrier region --
a realistic magnetoelectronic structure that can be grown epitaxially.

\begin{figure}
\centerline{
\includegraphics[width=5.8cm]{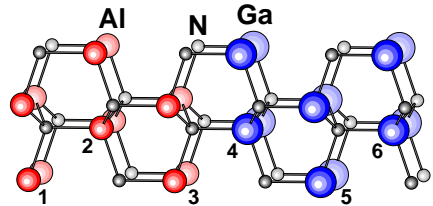}
}
\caption{Geometry of (2x2) AlN/GaN (0001) heterostructure: 
the large, intermediate and small spheres correspond to
Ga, Al and N atoms, respectively. Substitutional site locations 
of Cr atoms in the supercell denoted by numbers. }
\label{fig1}
\end{figure}

The AlN/GaN (0001) interface is modeled using (1x1), (2x2) and (2x4) 
supercells of pure or Cr-doped wurtzite GaN and AlN (0001) with 3 double layers 
of each material, cf. Fig. \ref{fig1}.
Note that a ``double layer'' is a single 
nitride layer, i.e., it consists of a group III element and nitrogen, so the (1x1), 
(2x2) and (2x4) supercells contain 12, 48 and 96 atoms, respectively. 
For the Al$_{1-x}$Ga$_x$N/GaN (0001) heterostructures, we used (2x2) supercells
with $x$=0.00, 0.25, 0.50, 0.75 and 1.00 in the layers, i.e.,
the content of Al (Ga) gradually decreases (increases) along the (0001) 
direction until a composition of pure GaN is formed.
The equilibrium relaxed geometry of the structures was determined 
via total energy and atomic forces minimization for 
the lattice parameters $a$ and $c$ and the internal parameter $u$.
Both the GaN and AlN regions of the heterostructure were allowed to relax 
with the same in-plane lattice constant \cite{Kim}, while the internal 
atomic relaxation for each lattice parameter $c$ provided different 
minimized III--N distances along the (0001) direction. 
Note, that during the optimization, {\it all} atoms were allowed to move 
in the x, y and z directions.

\hspace{-0.4cm}
\begin{minipage}{8.5cm}
\begin{table}
\caption{Calculated lattice parameters, \AA, 
and band gap values, eV, for bulk wurtzite AlN and GaN and 
for the AlN/GaN (0001) heterostructure. 
Experimental data are taken from Ref. 16.
}
\vspace{0.2cm}
\begin{tabular}{c|cccc|ccc}
& \multicolumn{4}{c|}{Lattice parameters} & \multicolumn{3}{c}{Band gap}  \\
        &  $a$  &  $c$ &$a_{exp}$&$c_{exp}$&  LDA & sX-LDA & Exp. \\ \hline
AlN     & 3.085 & 4.993 & 3.112 &  4.982  &  4.34 &  6.05  & 6.20 \\
GaN     & 3.142 & 5.194 & 3.189 &  5.185  &  2.00 &  3.35  & 3.39 \\
AlN/GaN & 3.127 &15.402 &       &         &  2.59 &  3.98  &      \\
\end{tabular}
\label{table1}
\end{table}
\end{minipage}

We employ the all-electron full-potential linearized augmented plane wave 
(FLAPW) method \cite{FLAPW} that has no shape approximation for the potential 
and charge density. Cut-offs of the plane-wave basis 
(16.0 Ry) and potential representation (81.0 Ry), and expansion in terms 
of spherical harmonics with $\ell \le$ 8 inside the muffin-tin spheres 
were used. Summations over the Brillouin zone were carried out 
using 32 special {\bf k} points in the irreducible wedge. 
In addition to the local density approximation (LDA) for the ground state
properties, we used 
the self-consistent screened-exchange LDA (sX-LDA) method \cite{sXLDA,mysx},
which is known to provide a considerably improved description of 
the excited state (optical) properties as compared to the LDA or generalized 
gradient approximation (GGA) calculations.

We first determine the effect of the interface on the structural and electronic 
properties of pure AlN/GaN (0001) and compare with those of bulk GaN and AlN.
Due to the lattice mismatch, 
both GaN and AlN are found to be mutually strained in the junction,
cf. Table \ref{table1}:
(i) The resulting in-plane lattice constant, 3.127 \AA, is 0.5 \% smaller 
(1.3 \% larger) than that of the bulk GaN (AlN) and is in agreement 
with the value observed \cite{Kim} for the thinnest AlN/GaN bilayer, 3.134 \AA.
(ii) The renormalized $c$ lattice constant of GaN in the heterostructure 
is larger than that of AlN by 0.195 \AA \, which is approximately equal
to the difference between the $c$ parameters of the bulk materials, cf.,
Table \ref{table1}. We also found that away from the junctions,
the Ga--N and Al--N distances along (0001) in each sub-unit tend 
to relax back towards their unstrained bulk values and 
differ by 0.7 \% (0.9 \%) from those in the bulk GaN (AlN). 
Further, the structural relaxation at the junctions affects the electronic 
properties, in particular, the optical bandgap of AlN/GaN (0001). 
In Table \ref{table1}, we present the bandgap values
of bulk GaN and AlN and the (1x1) AlN/GaN (0001) calculated
using both the LDA and sX-LDA methods. As expected, the LDA underestimates
the band gap of both bulk GaN and AlN by 30--40 \%, while the sX-LDA
gives a considerably improved description -- namely, a less than 2\%-difference 
from the experimental values. 
For the AlN/GaN (0001), we found that the valence band maximum and the conduction 
band minimum are formed by states of the GaN layers; thus, these states 
determine the calculated minimum band gap of the system to be 2.59 eV (3.98 eV) 
within the LDA (sX-LDA). In contrast, the states of the AlN layer located
further away from the junctions lie deeper in the valence and conduction bands 
giving a bandgap increase of $\sim$0.4 eV
within both the LDA and sX-LDA \cite{55}.

Next, we investigated the effect of the interface on 
the transition-metal impurity location and the magnetic properties of 
the doped structure. To this end, we performed calculations
of Cr doped substitutionally into the (2x2) AlN/GaN (0001). 
(For comparison of the electronic and magnetic properties, 
we also calculated Cr-doped bulk GaN and AlN.)
The choice of the (2x2) supercell is motivated by the recently calculated
critical separation between two Cr atoms in bulk GaN, $\sim$2.7 \AA, above 
which ferromagnetic (FM) coupling dominates the antiferromagnetic 
(AFM) one \cite{Das}. In our supercell case, the Cr-Cr distance is 6.3 \AA.
Now, to find the preferred site location of the magnetic impurity, 
we calculated the formation energies \cite{formation} of six relaxed 
structures with cation-substituted Cr, cf. Fig. \ref{fig1}.
The results, gathered in Table \ref{table2}, allow 
the following conclusions:

\hspace{-0.4cm}
\begin{minipage}{8.5cm}
\begin{table}
\caption{Comparison of the calculated formation energies, eV, 
and the magnetic moments on the Cr atoms and its tetrahedral N neighbors 
in plane (N$_{planar}$) and along the (0001) direction (N$_{apical}$), $\mu_B$, 
for the Cr-doped (2x2) AlN/GaN (0001) heterostructures. 
Site locations of the Cr atoms (in parentheses) correspond to those
in Fig. \ref{fig1}.}
\vspace{0.2cm}
\begin{tabular}{c|rrr|rrr} 
& \multicolumn{3}{c|}{AlN region} & \multicolumn{3}{c}{GaN region}  \\
              &  Cr(1) &  Cr(2) &  Cr(3) &  Cr(4) &  Cr(5)  &  Cr(6)  \\ \hline
$\Delta$E$_f$ & +1.825 & +1.817 & +1.827 &  0.000 &  +0.031 &  +0.010 \\
Cr            &  2.309 &  2.309 &  2.308 &  2.302 &  2.304  &  2.299  \\
N$_{planar}$  &$-$0.010&$-$0.011&$-$0.015&$-$0.020&$-$0.019 &$-$0.016  \\
N$_{apical}$  &$-$0.018&$-$0.007&$-$0.005&$-$0.003&$-$0.012 &$-$0.012  \\
\end{tabular}
\label{table2}
\end{table}
\end{minipage}

(i) The magnetic impurity segregates into the GaN region of the heterostructure
which serves as a ferromagnet (as shown below), while the AlN region is 
a non-magnetic part of the interface. The difference of the formation energies
for Cr in the AlN and in the GaN region of the heterostructure ($\sim$1.8 eV,
cf. Table \ref{table2}) shows little dependence on the distance from the junctions
\cite{55cathy} and agrees well with that for Cr-doped bulk GaN and AlN (2.0 eV).

(ii) Although the calculated local magnetic moments on the Cr atoms in the AlN/GaN, 
Table \ref{table2}, are similar to those obtained for 
the corresponding relaxed Cr-doped bulk GaN (2.304 $\mu_B$) and AlN (2.313 $\mu_B$), 
the presence of the junctions does affect the magnetic properties of the system. 
This can be seen from a comparison of the induced magnetic moments on 
the tetrahedral N neighbors of the Cr atom (i.e., three in-plane N and 
one apical N atoms, cf. Fig. \ref{fig1}), presented in Table \ref{table2}. 
We found that the magnetic moments are larger on the N atoms 
which bond with Ga than on those connected to Al atoms by 0.006 $\mu_B$ 
(or 0.009 $\mu_B$) for planar (or apical) N atoms, on average.
This finding can be explained by the fact that the states of the AlN layers
lie deeper in the valence band (as discussed above) and hence they are more 
screened as compared to the GaN layers \cite{mm}.
Therefore, at the junction where the in-plane N atoms bond with Ga and
the apical N atom bonds with Al, the induced in-plane spin polarization 
is dominant (case 4, Table \ref{table2}). In contrast, a higher
spin polarization along (0001) -- as compared to the in-plane magnetic moments 
-- is observed at the other junction (case 1, Table \ref{table2}).
Accordingly, we found the magnetic moments on planar and apical N atoms 
to be similar when Cr is equidistant from the junctions.

\begin{figure}
\includegraphics[width=3.0cm]{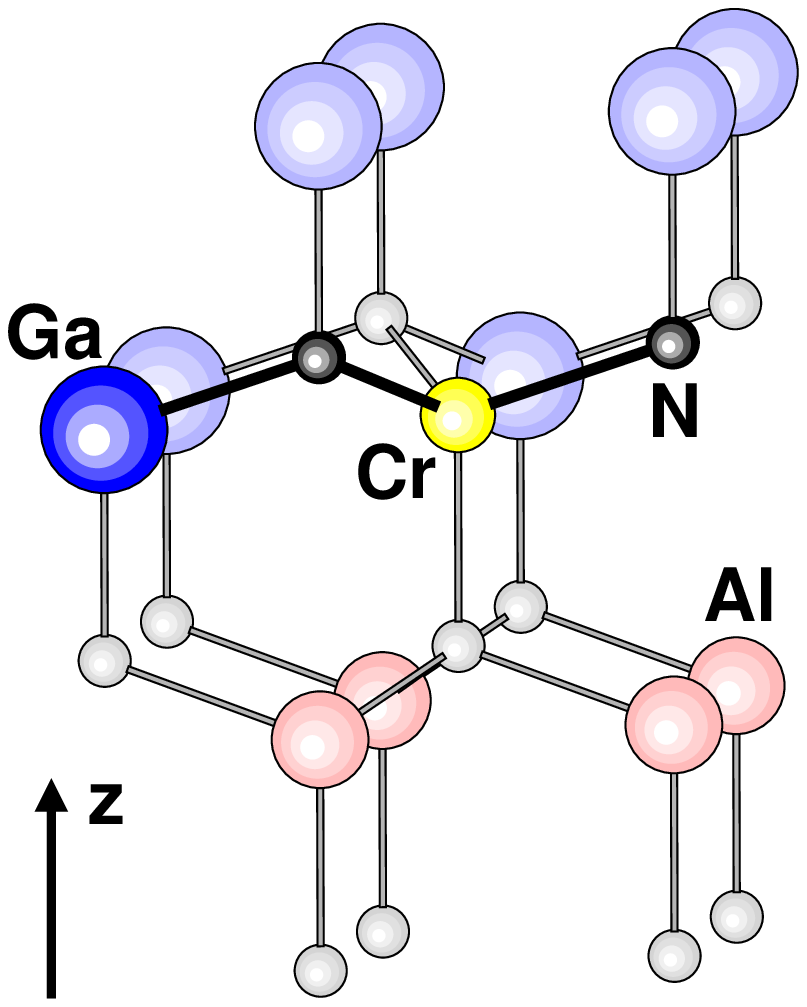}
\includegraphics[width=5.0cm]{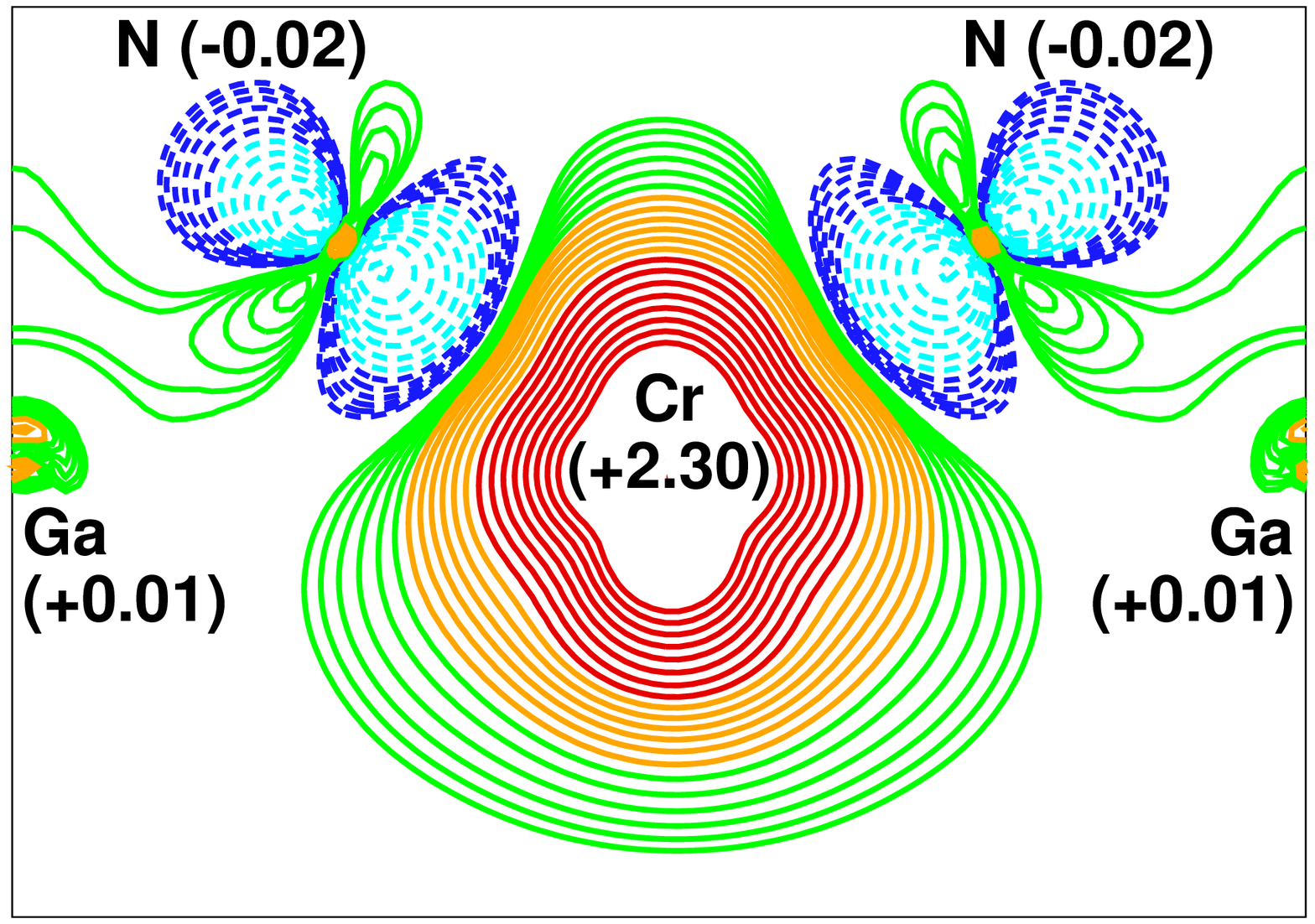}
\caption{Geometry of the most energetically favorable Cr site location
in the AlN/GaN (0001) heterostructure (left) and
contour plot of the calculated spin density distribution within a slice 
passing through this Cr atom and its planar N neighbors (right). 
Solid (dash) lines in the plot denote positive (negative) spin polarization. 
The calculated magnetic moment on the atoms, $\mu_B$, is given in parentheses. 
}
\label{fig2}
\end{figure}

A contour plot of the calculated spin density distribution within 
a slice passing through the Cr atom and its planar N neighbors 
(case 4 in Table \ref{table2}) is presented in Fig. \ref{fig2}.
In agreement with the itinerant $sp$-$d$ exchange model \cite{Kanamori},
the $p$-orbital of the N atoms which points toward the Cr atom 
has its magnetic moment antiparallel to the one on the magnetic impurity
giving rise to FM coupling between two Cr atoms. (Note, that the other 
two N $p$-orbitals possess positive spin polarization and 
create hopping channels for itinerant electrons.)
Indeed, from the (2x4) Cr-doped AlN/GaN (0001) supercell calculations,
we confirm strong FM coupling between two Cr atoms with an energy difference 
between FM and AFM orderings of 97, 106 and 110 meV/Cr atom for case 4, 5 and 6, 
respectively. Accordingly, the effective exchange interaction 
parameters -- estimated \cite{exchange} to be 22, 26 and 28 meV, respectively -- 
show a similar dependence on the Cr site locations,
which correlates with the decrease of the ratio between the induced 
magnetic moments on planar and apical N atoms, cf. Table \ref{table2}.

\begin{figure}
\includegraphics[width=4.1cm]{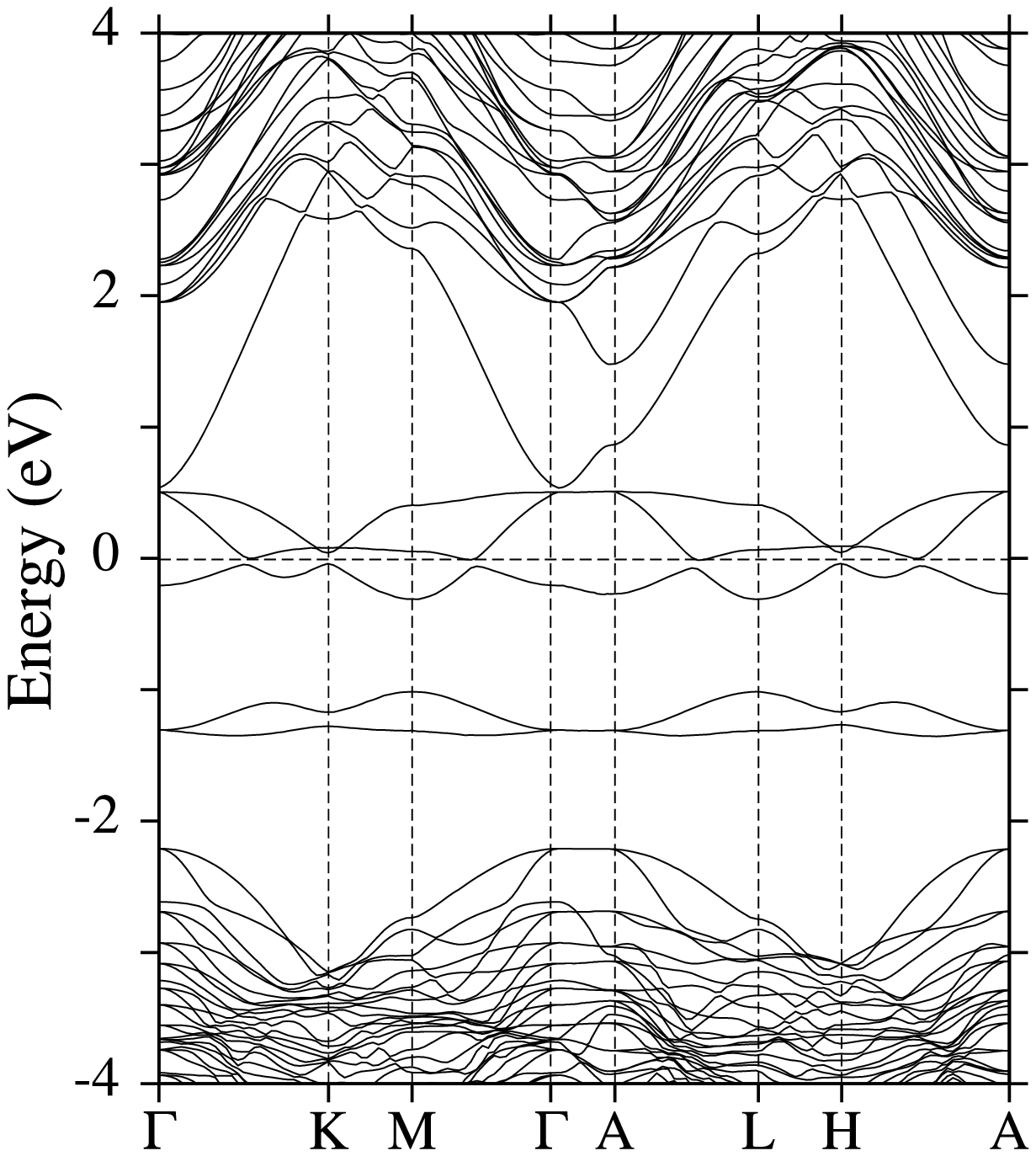}
\includegraphics[width=4.1cm]{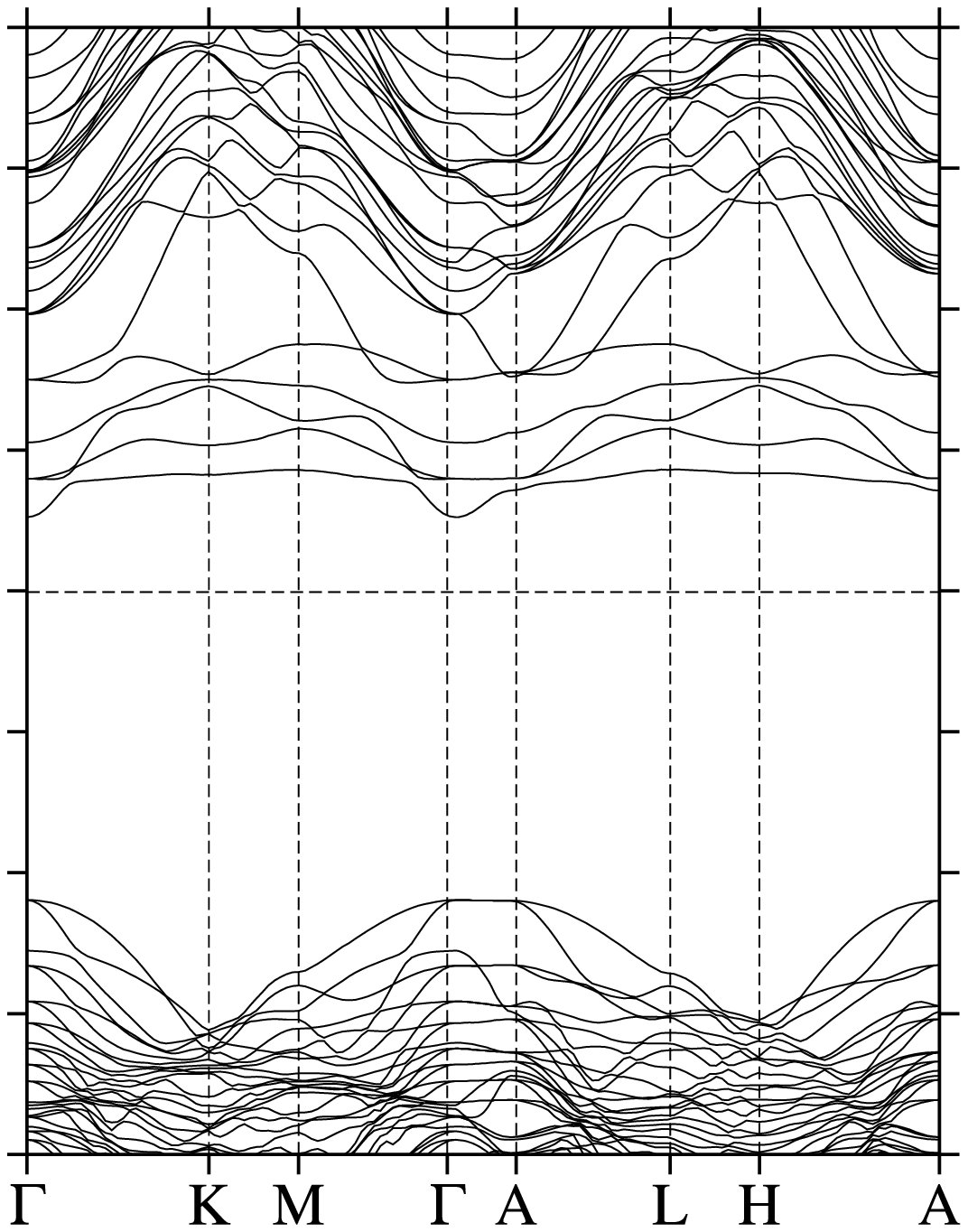}
\caption{Spin-resolved band structure along the high symmetry directions 
in the Brillouin zone for the most energetically favorable Cr-doped AlN/GaN 
(0001) heterostructure. The origin of the energy is taken 
at the Fermi level.
}
\label{fig3}
\end{figure}

(iii) Regardless of the Cr site location in the GaN region of the interface
(cf. cases 4-6 in Table \ref{table2} with a total energy difference 
of $\sim$$kT$), the three Cr-doped AlN/GaN systems share similar 
band structure features (the corresponding band offsets are discussed in 
a separate paper \cite{tobe}): The magnetic impurity $d$ states hybridized with 
the $p$ states of the neighboring N atoms form deep bands in the nitride bandgap, 
cf. Fig. \ref{fig3}. For the majority-spin channel, partially occupied triply 
degenerate $t_{2g}$ and fully occupied doubly degenerate $e_g$ bands are 
located at $\sim$2.5 eV and $\sim$1.1 eV, respectively, above the valence band maximum.
The exchange interaction splits the Cr $d$ states by $\sim$2 eV. These findings
are in agreement with previous studies \cite{Mryasov,Das} of Cr-doped bulk GaN.

Most significantly, we found that the Cr-doped AlN/GaN (0001) heterostructures
retain the desired half-metallic behavior (consistent with 
the calculated integer total magnetic moment of 3 $\mu_B$/cell)
with a bandgap in the spin minority channel of $\sim$2.7 eV, cf. Fig. \ref{fig3}.
This leads to a complete, i.e., 100 \%, spin polarization of the conduction 
electrons and thus makes the system attractive for high-efficiency magneto-electronic 
devices -- specifically, magnetoresistive tunnel junctions.

\begin{figure}
\includegraphics[width=7.6cm]{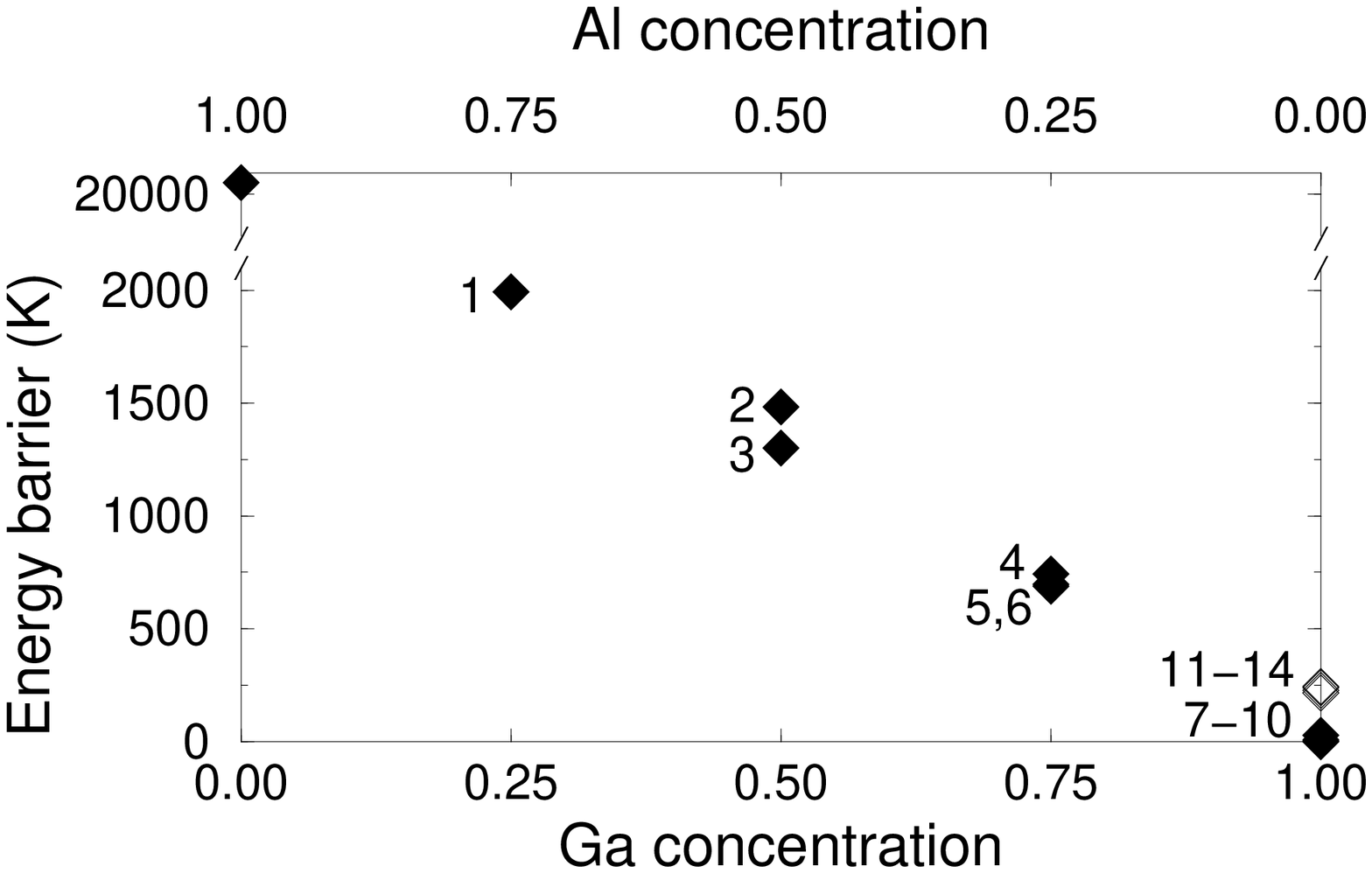}
\includegraphics[width=7.6cm]{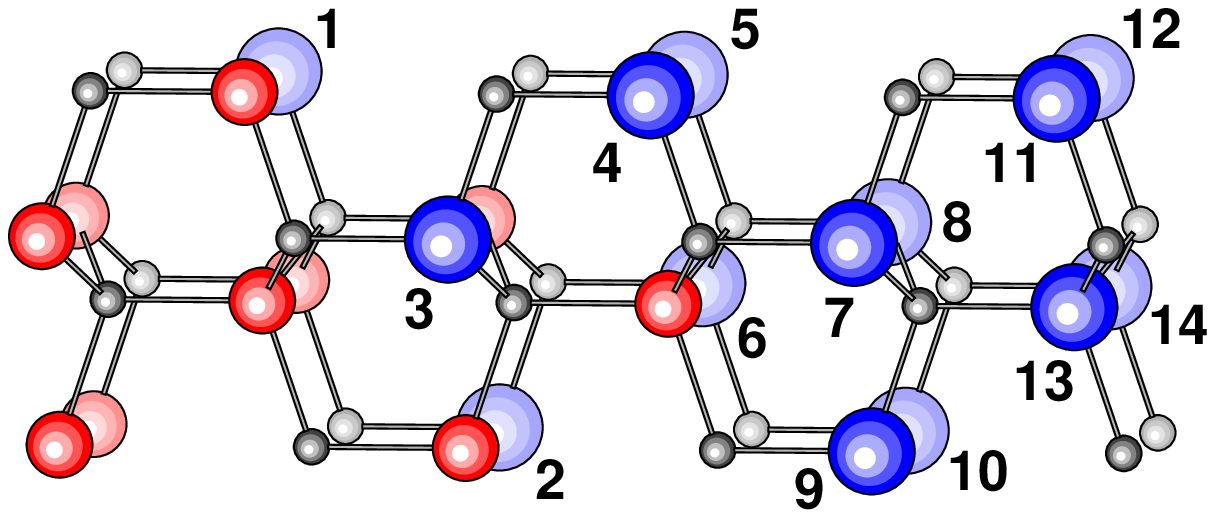}
\caption{Calculated relative formation energies, K, as a function of the 
cation concentration $x$ in the Cr-doped graded Al$_{1-x}$Ga$_x$N/GaN (0001) 
heterostructures and geometry of the corresponding supercell. The numbers
on the graph correspond to those in the supercell and denote
the Ga site locations substituted with a Cr atom. 
}
\label{fig4}
\end{figure}

Based on these results, in particular on the preference of Cr atoms 
to substitute Ga rather than Al atoms in the AlN/GaN (0001), 
we model a realistic (in the sense of practical applications) magneto-electronic 
system using (2x2) supercells of Cr-doped graded Al$_{1-x}$Ga$_x$N/GaN (0001) 
heterostructures.
From a comparison of the total energies for the relaxed structures with 
different Cr site locations, we found that the magnetic impurity prefers 
to substitute the Ga atoms which are located in the pure GaN (0001) layers 
(cf. cases 7-14 in Fig. \ref{fig4}). A significant total energy increase 
is obtained for the structures where Cr substitutes the Ga atoms 
in the mixed nitride (0001) layers (cf. cases 1-6 in Fig. \ref{fig4}).
This energy dependence on the composition suggests the possibility to control 
the width and height of the energy barrier by adjusting 
the Ga/Al ratio in the layers perpendicular to the growth direction -- 
an important ingredient for device design optimization.

In summary, based on the first-principles calculations, we predict that 
efficient spin injection can be achieved using a ferromagnetic
GaN:Cr electrode in conjunction with an AlN tunnel barrier.
Since the Cr-doped AlN/GaN (0001) heterostructures are found to be 
half-metallic, one can expect a pronounced increase of magnetoresistance 
in the nitride-based magnetic tunnel junctions.
These findings make the system highly attractive for spintronic applications
and call for experimental investigations.

Work supported by DARPA (grant 02-092-1/N00014-02-1-05918).
Computational resources provided by the NSF supported 
National Center for Supercomputing Applications, Urbana-Champaign.

\end{multicols}

\begin{thebibliography}{99}
\bibitem{mysx}
We used cut-off parameters of 12.96 Ry 
in the wave vectors and $\ell \le$ 3 inside the muffin-tin spheres.
\bibitem{55}
From additional calculations for (1x1) AlN/GaN (0001) with 5 double layers 
of each material, we found that the difference between the bandgap values
for the GaN layers and the AlN layers located further away from the junctions
is 0.4 eV (1.3 eV) within the LDA (sX-LDA).
\bibitem{formation}
We assume metal-rich growth conditions, i.e., the atom removed from (added to) 
the AlN/GaN is incorporated into (comes from) the corresponding bulk reservoir;
thus, the formation energy is calculated as 
E$_f$ = E$_{int:Cr}$ -- E$_{int}$ -- E$_{Ga(Al)}$ + E$_{Cr}$,
where E$_{int:Cr}$, E$_{int}$, E$_{Ga(Al)}$, and E$_{Cr}$ are
the total energy of the Cr-doped interface, the corresponding undoped interface, 
bulk orthorhombic Ga (fcc Al) and bcc Cr, respectively.
\bibitem{55cathy}
From additional calculations of Cr-doped (2x2) AlN/GaN (0001) with 5 double 
layers of each material, we found a negligible variation in the formation 
energies ($<$0.04 eV) across the heterostructure.
\bibitem{mm}
We also found that the Cr induced magnetic moments on its neighboring Ga atoms
($\sim$0.014 $\mu_B$) are larger than those on the Al atoms ($\sim$0.010 $\mu_B$).
\end{thebibliography}
\end{document}